\documentclass[pra,twocolumn,showpacs,floatfix,titlepage,superscriptaddress]{revtex4-1}

\usepackage{bm}
\usepackage{graphics} 
\usepackage{graphicx}  
\usepackage{latexsym}                % to get LASY symbols
\usepackage{amsmath}     
\usepackage{amsfonts}  
\usepackage{amssymb}
\usepackage{amsthm}
\usepackage{dcolumn}
\usepackage{color}
\usepackage{mathrsfs}
\usepackage[left]{lineno}

\renewcommand{\phi}{\varphi}
\renewcommand{\>}{\right \rangle}

\newcommand{\ket}[1]{\left |#1\>}

\newcommand{\be}{\begin{equation}}
\newcommand{\ee}{\end{equation}}
\newcommand{\bea}{\begin{eqnarray}}
\newcommand{\eea}{\end{eqnarray}}

\begin{document}

\title{Analysis and minimization of bending losses in discrete quantum networks}

\author{G.~M.~Nikolopoulos}
\affiliation{Institute of Electronic Structure \& Laser, FORTH, P.O.Box 1527, GR-71110 Heraklion, Greece}

\author{A. Hoskovec}
\affiliation{Department of Physics, FNSPE, Czech Technical University in Prague, B\v rehov\'a 7, 115 19 Praha 1, Star\'e M\v{e}sto, Czech Republic}

\author{I. Jex}
\affiliation{Department of Physics, FNSPE, Czech Technical University in Prague, B\v rehov\'a 7, 115 19 Praha 1, Star\'e M\v{e}sto, Czech Republic}

\date{\today}

\begin{abstract}
We study theoretically the transfer of quantum information along bends in two-dimensional discrete lattices.  
Our analysis shows that the fidelity of the transfer decreases considerably, as a result of interactions in the neighbourhood of the  bend. It is also demonstrated that such losses can be controlled efficiently 
by the inclusion of a defect.  
The present results are of relevance to various physical implementations of quantum networks, where 
geometric imperfections with finite spatial extent may arise as a result of bending, residual stress, etc.  
\end{abstract}


\maketitle

%%%%%%%%%%%%%%%%
\section{Introduction}
%%%%%%%%%%%%%%%%

Studies on the faithful transfer of quantum information and the engineering of discrete quantum networks 
have been focused mainly on one-dimensional topologies, and the transfer of signals between the two 
ends of a quantum chain \cite{review}. 
However, in analogy to conventional networks \cite{book1}, the prospect of large-scale quantum information processing (QIP) and networking, 
irrespective of the physical platform, require efficient complex signal manipulations (such as routing, splitting, switching, etc), which are possible only in higher-dimensional geometric arrangements. This necessity has motivated studies on state transfer in various 2D arrangements \cite{2Dbib}, most of which rely on nearest-neighbour (NN) Hamiltonians. 

Bends are expected to be at the core of any 2D configuration, yet their role on the fidelity of the transfer has not been investigated in the literature so far.  
We address this issue by investigating the transfer of quantum signals 
along a bent quantum chain whose operation is based on two different faithful-communication (FC) Hamiltonians 
with NN couplings \cite{prot2,UniProt}, one of which has been employed in many of the 2D arrangements of \cite{2Dbib}. 
By construction, these Hamiltonians ensure the faithful communication through the unbent chain, whereas their performance in the presence of bends is not known, and will be discussed here.  
In many physical realizations, the coupling between two adjacent sites is directly related to their spatial separation (e.g., see \cite{BeyondNN,HamEng,exp}). 
Hence, interactions beyond NNs are expected to get enhanced in the neighbourhood of bends, disturbing the communication. 
One way to circumvent such problems is to engineer new FC Hamiltonians for the bent chain taking into account interactions beyond NNs 
\cite{BeyondNN}; a rather tedious process for long chains. Here, we adopt another approach namely,  
the minimization of  bending losses on the basis of the unperturbed NN Hamiltonians by means of minimal external control (i.e., without elaborate sequences of pulses and measurements). To this end, a thorough analysis on the bending losses has been performed. 

In the following section we formulate the problem, whereas in Sec. III we analyse the bending losses and provide a way for their minimization. We conclude with a discussion in Sec. IV. 

%%%%%%%%%%%%%%%
\section{Formalism}
%%%%%%%%%%%%%%%

The 2D arrangement under consideration pertains to $N$ identical sites and is depicted in Fig. \ref{fig1}. 
Each site is associated with a qubit and the entire structure operates as a bent chain of qubits that 
interact according to a Hamiltonian of the form 
\begin{subequations}
\label{Ham_full}
\bea
\hat {\mathscr{ H}}=\hat{{\mathscr H}}_{0} +\hat{{\mathscr V}}(\theta),
\label{ham}
\eea 
where $\hat{{\mathscr H}}_{0} $ is the unperturbed Hamiltonian corresponding to the  unbent wire i.e., to $\theta=0$, 
and $\hat{\mathscr V}(\theta)$ is the perturbation associated with the bending. We consider two protocols, which ensure the faithful transfer of information between the two ends of 
the unbent chain, and our task is to analyse their robustness against perturbations that stem from the bending of the chain.  
Both of the protocols pertain to centrosymmetric channels with NN interactions and the unperturbed FC Hamiltonian is of the form  \cite{remark9}
\bea
\hat{{\mathscr H}}_{0} = \sum_{j=1}^{N} \varepsilon_j \hat{a}^\dag_{j} \hat{a}_{j} 
+\sum_{j=1}^{N-1}\Omega_{j,j+1}
(\hat{a}^\dag_{j} \hat{a}_{j+1}+\hat{a}^\dag_{j+1} \hat{a}_{j}),
\label{PST_ham1}
\eea
where $\hat{a}^\dag_{j}$ is the creation operator for an excitation on the  
$j$th site of the channel with energy $\varepsilon_j$, 
and $\Omega_{i,j}$ is the coupling between the sites with indices $i$ and $j$. 

{\em Protocol 1.} The first protocol is characterized by 
$\varepsilon_j=\varepsilon$, $\Omega_{j,j+1}=\Omega_0\,\,\forall j\neq 1,N-1$, and 
$\Omega_{1,2}=\Omega_{N-1,N}=\Omega$ \cite{UniProt}. For a given $N$, the ratio $\Omega/\Omega_0$ 
is chosen so that faithful transfer of information between the two ends of the chain occurs  
at time $T_1^{(0)}$ \cite{remark10}. 

{\em Protocol 2.} The second protocol also pertains to 
resonant sites, while the couplings along the entire chain are engineered according to  
$\Omega_{j,j+1}=\Omega_0 \sqrt{(N-j)j}$ \cite{prot2}. 
In contrast to protocol 1, this scheme promises ideally perfect transfer at time $T_2^{(0)}=\pi/(2\Omega_0)$. 
From now on, we also refer to a chain that operates in protocol $i$ as chain (channel)  $i$. 

\begin{figure}
\centerline{\includegraphics[scale=0.4]{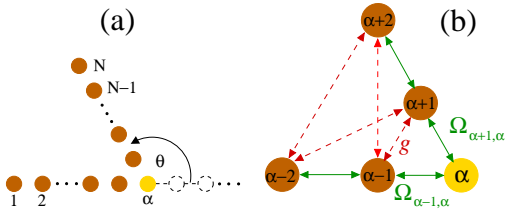}}
\caption{(Color online) 
(a) A bent quantum chain of $N$ sites.   (b) A closeup of the bend. Green (solid) arrows denote NN interactions of strength $\Omega_{i,j}$. Red (dashed) arrows denote interactions beyond NNs, with the strongest one (of  strength $g$) corresponding to the first neighbours of the corner site $\alpha$.}
\label{fig1}
\end{figure}

In various implementations of quantum networks, the coupling constant $\Omega_{i,j}$ depends, among other parameters, on the inter-site separation $r_{i,j}$; typically, it is expected to increase as we decrease $r_{i,j}$ (e.g., see \cite{BeyondNN,HamEng,exp}). 
The details of the dependence of $\Omega_{i,j}$ on $r_{i,j}$  may vary from realization to realization, but also from site to site within the same network because of  disorder and 
imperfections. As long as, however, such a type of disturbances are sufficiently weak (a necessary requirement for QIP) 
it is reasonable to assume that, to a good approximation the spatial dependence of $\Omega_{i,j}$ is governed 
by a law that is universal for the entire network in a particular realization. 

Let $\alpha\in [2,N-1]$ denote the index for the corner site at the bend.  
The separation between the non-neighbouring sites with indices $\alpha+k$ and $\alpha-k^\prime$ (where $k,k^\prime=1,2,\ldots$) decreases with increasing $\theta$, while the separation between NNs is assumed to remain unaffected (see Fig. \ref{fig1}). 
Hence, couplings  between non-neighbouring sites are expected to be present for sufficiently sharp bends, and have to be taken into account. 
In an attempt to understand the effect of the bending on chains 1 and 2, throughout this work 
we focus on a regime of $\theta$ where the perturbation is dominated by the coupling between the first-order neighbours of the corner site with indices 
$\alpha\pm 1$, while  the effect of couplings between higher-order neighbours on the time scales of interest can be neglected. Hence, the perturbation in Eq. (\ref{Ham_full}) can be chosen as   
\bea
\hat{{\mathscr V}}(\theta) = g(\theta)(\hat{a}^\dag_{\alpha-1} \hat{a}_{\alpha+1}+\hat{a}^\dag_{\alpha+1} \hat{a}_{\alpha-1}).
\label{Vham}
\eea 
\end{subequations}
The regime of $\theta$ for which such an assumption is valid,  is intimately connected to the details of a given implementation. In a more general context, the perturbation (\ref{Vham}) is expected to describe other types of undesirable geometric spatially-localized imperfections associated  
e.g., with residual stresses \cite{remark1}, in a particular realization of the chains. In either case, the key point is that the dynamics of the chains, are expected to be determined by the strength of the perturbation $g$, relative to the NN couplings around the corner, rather than the actual origin of the perturbation.  
In the limit of weak perturbation ($g\ll \Omega_{\alpha,\alpha\mp 1}$) the performance of the bent channel is expected to be rather close to the
performance of the unbent one, whereas large deviations are expected for $g\geq \Omega_{\alpha,\alpha\mp 1}$. 
In view of the absence of related experiments, we keep the following analysis  
in a rather general theoretical framework, by introducing the ratio $\kappa=g/\Omega_{\max}$ \cite{remark2}, where $\Omega_{\max}\equiv\max\{\Omega_{i,j}\}$ \cite{remark5}. 
  
\begin{figure}
 \centering
\includegraphics[width=8.cm,clip]{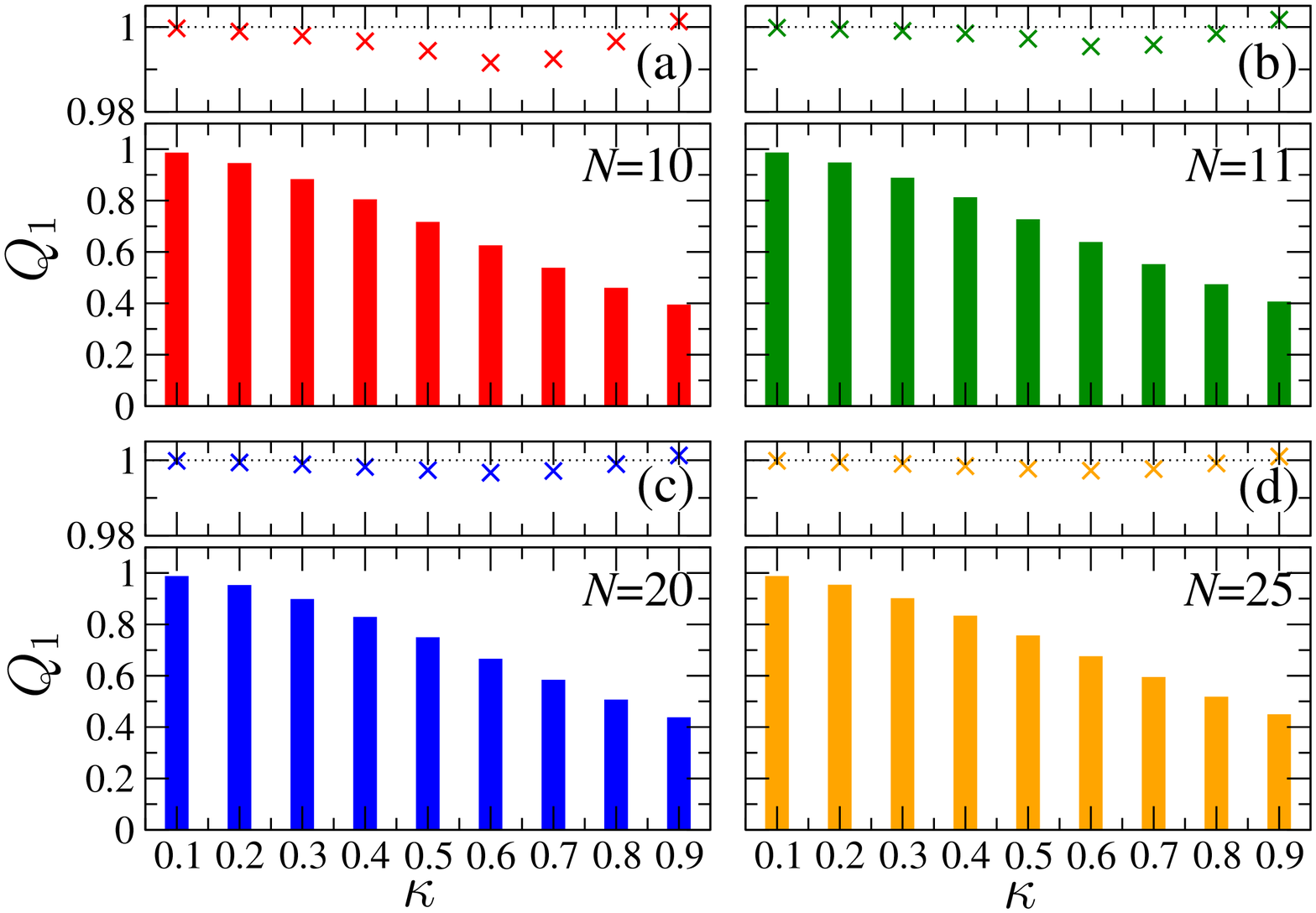}
\hfill
\includegraphics[width=8.cm,clip]{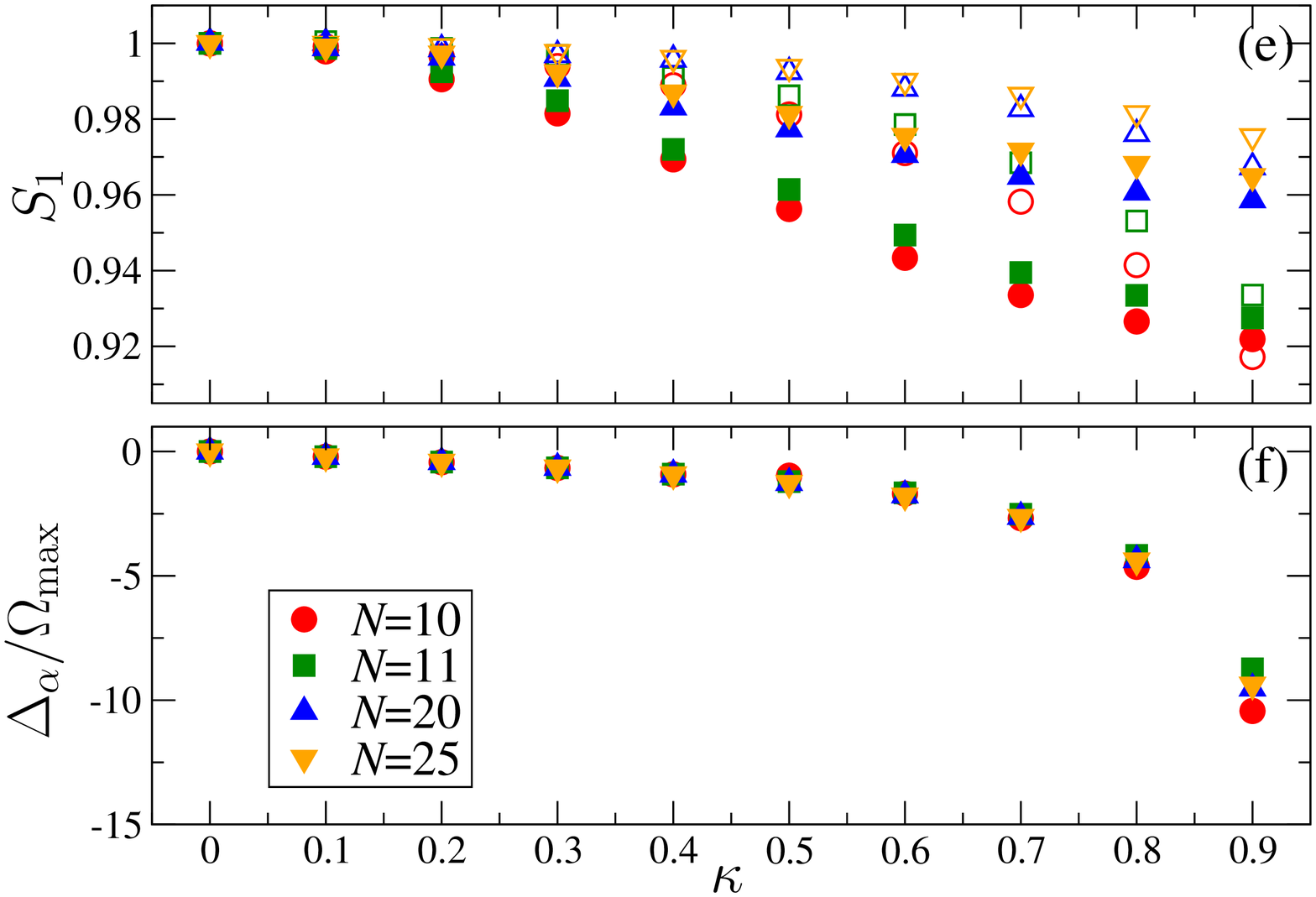}
\caption{(Color online) Performance of the bent chain 1 for various $N$. 
(a-d) The normalized probability of transfer $Q_i$, for the bent chain 
without (bars), and with ($\times$) corner defect.  
Corner sites:  
(a,b) $\alpha=6$;  (c) $\alpha=11$; (d) $\alpha=13$. 
(e) Time of transfer through the bent chain without (open symbols) and with (filed symbols) corner defect. 
(f) Optimal detunings of the defect corner site relative to the other sites \cite{remark4}.}
\label{fig2}
\end{figure}  
  
Following \cite{review,prot2,UniProt,BeyondNN,HamEng}, the chain is initially prepared in an eigenstate of $\hat{\mathscr H}$, and the information to be transferred  
is encoded in the state of the first qubit. The Hamiltonian (\ref{Ham_full}) preserves the number of excitations  
and thus, the system is restricted to the one-excitation Hilbert space 
throughout its evolution. Various degrees of freedom that may be associated with the quantum state 
of the information carrier are assumed to be preserved on the time scales of interest, and thus the problem 
of the state transfer boils down to the transfer of the excitation. 
The computational basis can be chosen as $\{\ket{j}\}$, where $\ket{j}$ 
is the state with one excitation on the $j$th site. 
In the Schr\"odinger picture the state of the system at any  time $t$ is given by 
$\ket{\Psi(t)}=\sum_j A_j\ket{j}$, where $|A_j(t)|^2$ is the probability for the excitation to occupy the 
$j$th site at time $t$. Initially, $\ket{\Psi(0)}=\ket{1}$, and the evolution of the amplitude $A_j$ for protocol 
$i$ and given $\alpha$ is governed by 
\bea
{\rm i}\frac{\mathrm{d}A_{j}(t)}{\mathrm{d}t}&=&\varepsilon_{j}A_{j}+\Omega_{j-1,j}A_{j-1}+\Omega_{j,j+1}A_{j+1}\nonumber\\ 
&&+g \delta_{j,\alpha+1} A_{\alpha-1}+g\delta_{j,\alpha-1} A_{\alpha+1}, 
\label{eom}
\eea
with $j\in[1,N]$ and $\delta_{m,n}$ is the Kronecker's delta. 
The last two terms are associated with the bend, and they were not present in any of the previous investigations on the 
two protocols \cite{2Dbib,prot2,UniProt}. We are interested in the probability for the excitation to occupy the 
$N$th site at time $t$ when protocol $i$ is used,  which is given by $P_i(t)\equiv|A_N(t)|^2$.  
The corresponding probabilities for the unbent chains are denoted by $P_{i}^{(0)}(t)$. 

\begin{figure}
 \centering
\includegraphics[width=8.cm,clip]{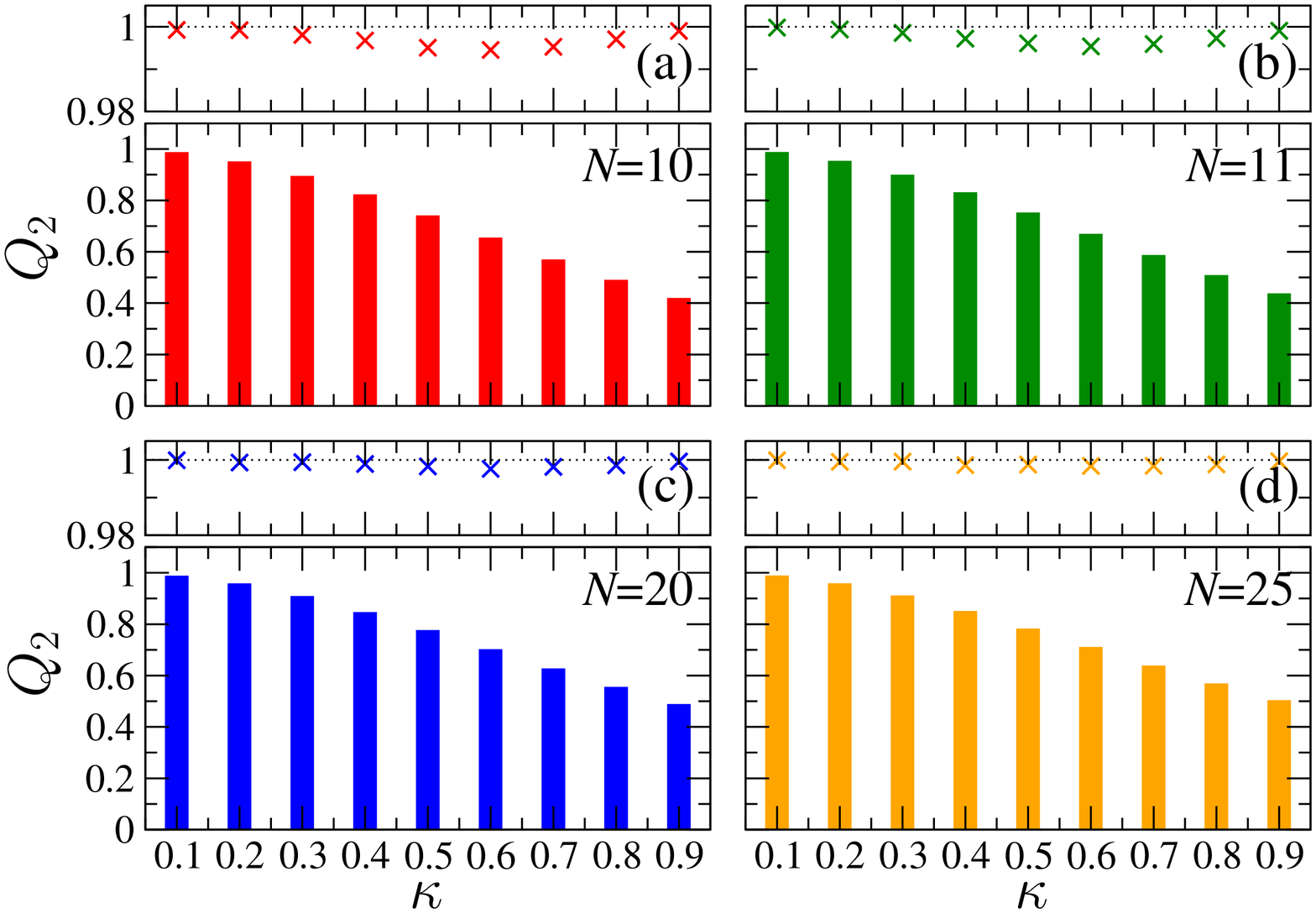}
\hfill
\includegraphics[width=8.cm,clip]{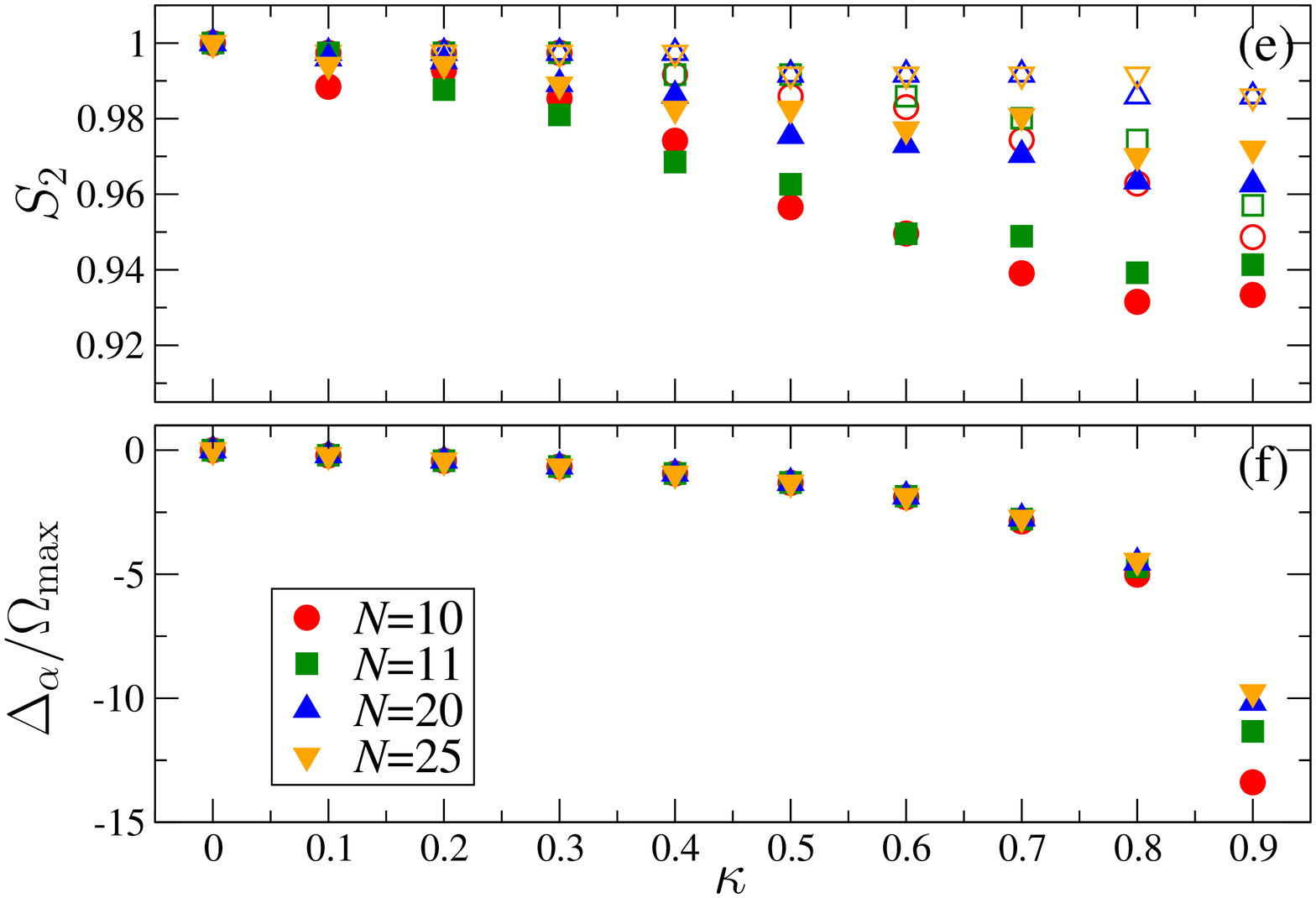}
\caption{(Color online) As in Fig. \ref{fig2} for chain 2. }
\label{fig3}
\end{figure}

%%%%%%%%%%%%%%%%%%%%%%%%%%%%
\section{Simulations}
%%%%%%%%%%%%%%%%%%%%%%%%%%%%

Equations (\ref{eom}) were solved numerically for various values of $\alpha$ and $\kappa\in[0,1]$, keeping track of the first maximum of $P_i(t)$ and the corresponding time $T_i$, at which  this is attained \cite{remark6}. 
The evolution was restricted to times $t\in[0,T_i^{(0)}]$, since the perturbation accelerates the transfer relative to the unbent channels (intuitively speaking, the coupling strength around the corner increases). Given that both of $P_1^{(0)}$ and $T_1^{(0)}$ are functions of $N$ \cite{UniProt}, for the sake of 
comparison our results are presented in terms of the ratios $Q_i\equiv P_i/P_i^{(0)}$ and $S_i\equiv T_i/T_i^{(0)}$. 
Most of the results presented here pertain to a bend in the middle of the chains. Analogous observations and conclusions hold for all $\alpha\in [2, N-1]$, and thus we do not show related plots \cite{remark3a}.

%%%%%%%%%%%%%%%%%%%%%%%%%%%%%%
\subsection{Analysis of bending losses} 
%%%%%%%%%%%%%%%%%%%%%%%%%%%%%%

According to the histograms of Figs. \ref{fig2} and \ref{fig3}, for both protocols $Q_i\approx 1$ for relatively weak perturbations 
(up to $\kappa\approx 0.2$ ), and drops as we increase $\kappa$. For a given protocol, there do not seem to exist major differences 
between even and odd values of $N$, whereas the decrease becomes slightly faster as we decrease $N$. This is due to the spatial extent of 
the perturbation, which pertains to two sites around the corner site. The fraction of perturbed over unperturbed sites is thus getting smaller as 
we increase $N$, and the effect of the perturbation becomes less significant for a given value of $\kappa$.
A Gaussian fit turns out to be a rather good approximation for the estimated points $Q_i(\kappa)$, 
and the ratio of the corresponding widths for the two protocols (associated with the decrease of $Q_i$ for increasing $\kappa$), is estimated to about $1.08$; a fact that shows how close the responses of the two protocols to the perturbation are.

The effect of the bend on the time of transfer for various $N$ is depicted in Figs. \ref{fig2}(e) and \ref{fig3}(e) (open symbols only).  For both protocols the transfer is accelerated relative to the unbent chains, 
with the acceleration being slightly more pronounced for chain 1. For a given protocol, the acceleration becomes less pronounced as we increase $N$; a behaviour that can be attributed again to the finite spatial extent of the perturbation. Moreover, the bend seems to affect mostly the fidelity  and secondly the time of the transfer. 

\begin{figure}
 \centering
\includegraphics[width=8.cm,clip]{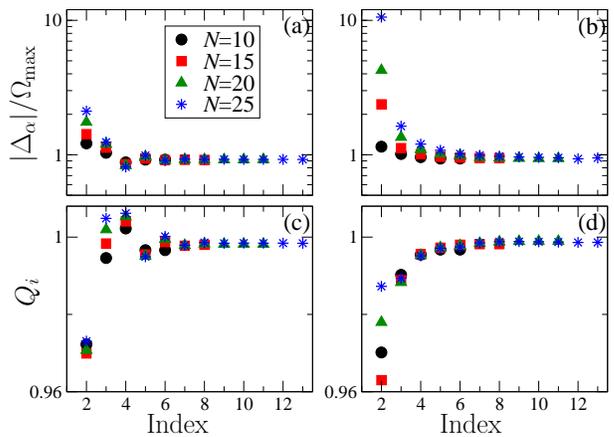}
\caption{(Color online) Performance of  bent chains for different positions of  the corner, and for fixed  $\kappa=0.4$. Different points for each $N$ are obtained by increasing $\alpha$ from  $2$ to 
$\lceil N/2\rceil$ with step 1 \cite{remark3b}.  (a,c) Protocol 1 and (b,d) Protocol 2. }
\label{fig4}
\end{figure}

%%%%%%%%%%%%%%%%%%%%%%%%%%%%%%%%%%
\subsection{Minimization of bending losses}
%%%%%%%%%%%%%%%%%%%%%%%%%%%%%%%%%%
Various approaches for the minimization of the bending losses have been employed. The most efficient we found pertains to the introduction of a defect by adjusting the energy of the corner site, while keeping all the other parameters the same. 
Let us denote by $\Delta_\alpha$, the detuning of the corner site relative to the other sites of the chains. This detuning is optimised so that the transfer 
from the first to the last site is maximized for a given strength of the perturbation $\kappa$ and for  times $t\in[0,T_i^{(0)}]$. 
The optimal values of $\Delta_\alpha$ for the two protocols and various $N$ are depicted in Figs. \ref{fig2}(f) and \ref{fig3}(f) \cite{remark4}, 
while the corresponding values of $Q_i$ are depicted with symbols $(\times)$ in Figs. \ref{fig2}(a-d) and \ref{fig3}(a-d). Clearly,  for given $N$, $\kappa$ and $\alpha$,  
there is an optimal value of $\Delta_\alpha$ for which the probability of transfer is above $99\%$ of the corresponding probability for the unbent chains \cite{remark8}.  
The optimal detuning turns out to be 
negative for all the tested parameters and increases (in absolute value) as we increase $\kappa$. Surprisingly enough, for both protocols, the optimal values of $\Delta_\alpha$ for various $N$ do not differ substantially throughout the entire regime of $\kappa$. Furthermore, to a good approximation, 
for both protocols and for all $N$, the behaviour of $\Delta_\alpha$ for $\kappa\lesssim 0.7$ is linear, with estimated slopes $-2.361$ and $-2.573$ 
for protocols 1 and 2, respectively.  
In contrast to $Q_i$,  the effect of the optimization on $S_i$ is not so significant  (filled symbols in Figs. \ref{fig2}(e) and \ref{fig3}(e)); overall, the transfer is a bit faster 
relative to the perturbed chains without optimization (open symbols). 

\begin{figure}
\centering
\includegraphics[width=8.cm,clip]{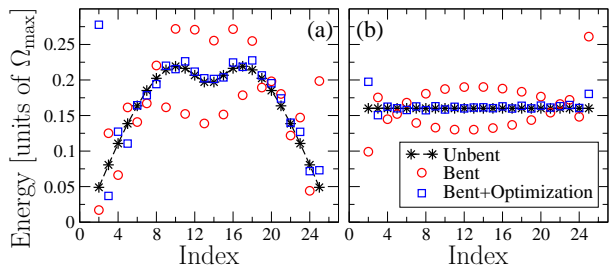}
\caption{(Color online) Difference between successive eigenenergies of the Hamiltonians for $N=25$ and  $\alpha=12$. (a) Protocol 1, $\kappa=0.5$; (b) Protocol 2, $\kappa=0.3$.}
\label{fig5}
\end{figure}

According to Figs. \ref{fig4}(a,b) the optimization is equally expensive for the two chains, in the sense that detunings of the same order are required in order to minimize the losses against the same disturbance $\kappa$. It becomes, however, particularly expensive when the bend is close to the ends, and in this respect such arrangements should be avoided. The optimization seems to work efficiently irrespective of the position of the bend on the chains.

A close inspection of  Figs. \ref{fig2}(e,f), \ref{fig3}(e,f), and \ref{fig4}(a,b) reveals that in general the optimal detunings are not so large to allow for adiabatic elimination of the corner site, and thus the reduction of the $N$-site bent chain to an effective $(N-1)-$site chain with NN couplings only. Hence, to gain further insight into the role of the bends and the minimization of the associated losses, 
we have analysed  the spectrum of the Hamiltonians (one-excitation sector) for the various cases \cite{remark7}. In Fig. \ref{fig5} we plot the separation between successive eigenvalues for the unperturbed Hamiltonian as well as the perturbed Hamiltonian with and without optimisation. Clearly, the presence of the perturbation affects significantly the relative  position of the eigenvalues, and this disturbance is responsible for the observed decrease of the probability of transfer. When the detuning of the corner site is optimized, however, the initial distribution is restored to a large extent, minimizing thus the losses. Some deviations at the borders are not of great importance since the overlap of the initial state $\ket{\Psi(0)}$ with the corresponding eigenvectors is negligible (i.e., the overlap is peaked around the center).
   
%%%%%%%%%%%%%%%%%%%%%%%%
\section{Discussion}
%%%%%%%%%%%%%%%%%%%%%%%%

We have discussed the transfer of quantum information along bent quantum chains that operate according to known FC Hamiltonians with NN interactions. Bends are at the core of various 2D configurations that have been discussed in the literature \cite{2Dbib}, yet their effects on  the transfer have been neglected, and the related investigations were focused on NN interaction Hamiltonians. Our analysis shows that the transfer is distorted significantly by  interactions beyond nearest-neighbours that stem from the bend. Nevertheless, the limited spatial extent of the perturbation, allowed us to minimize efficiently such losses by controlling the energy  of the corner site.  Large-scale QIP requires reliable and efficient navigation of quantum signals in higher-dimensional networks, where bends are expected to play a pivotal role. Our work sheds light on the role of such bends facilitating the engineering of reliable quantum networks in higher dimensions, including the 2D geometric arrangements of \cite{2Dbib}. 

The present results have been obtained in a rather general theoretical framework, and are expected to be of relevance to various physical implementations of quantum networks pertaining e.g.,  to quantum dots, optical lattices or photonic lattices \cite{prot2,UniProt,exp,Chr01}. In contrast to analogous theoretical and experimental studies in the context of discrete soliton networks (see \cite{Chr01} and references therein), our work pertains to linear networks whereas not all of the couplings between adjacent sites are the same. The  versatility of photonic lattices, however, allows for engineering of various coupling configurations and geometric arrangements with bends \cite{Chr01}, so that  our main observations can be confirmed experimentally in this context, with today's technology. 

In the weak-coupling regime the coupling between the waveguides with indices $j$ and $(j+1)$ depends exponentially on their separation $d_{j,j+1}$ i.e., we have $C_{j,j+ 1} = \eta \exp \left(-\xi d_{j,j+ 1}\right),$
where $\eta,\xi$ are open parameters to be determined by fitting to related experimental data for a particular setup \cite{exp}. 
Given this exponential law, one can readily show that a particular coupling distribution $\{\Omega_{j,j+1}\}$ is obtained in practise if the distance 
between successive waveguides is chosen according to $d_{j,j+1}=-\ln\left(\eta^{-1}\Omega_{j,j+1}\right )\xi^{-1}$. 
As typical values for the open parameters at the wavelength $\lambda=633$ nm, we may consider $\eta\sim 19.5$ cm$^{-1}$ and $\xi=0.152\,\mu$m$^{-1}$. Hence, the separations required e.g., for the realization of the  coupling distribution of protocol 2 in a lattice of $N=9$ waveguides of length $L=10$ cm, range from about 21.86 to 24.88 $\mu m$. Such separations are well within reach of current technology used in fabricating photonic lattices \cite{exp}. In view of the above exponential law, one can also readily show that NN couplings are almost two orders of magnitude larger than the couplings beyond nearest neighbours. One can thus fabricate higher-dimensional array configurations with NN couplings, but in the neighbourhood of bends interactions beyond NN set in locally \cite{Chr01}.   In the context of waveguides, the detuning of the corner site relative to the others can be achieved by altering the core refractive index or the core radius of the corner waveguide along the lines of \cite{Chr01}.  Such an adjustment has been demonstrated and studied experimentally for bent photonic arrays recently \cite{Hein11}. 
As mentioned before, however, these studies pertain to discrete soliton networks, whereas our work paves the way for manufacturing reliable  higher-dimensional linear networks with engineered couplings, by minimizing losses associated with bends.

In closing, we would like to emphasize that in principle there are infinitely may state-transfer Hamiltonians \cite{HamEng},  
and the present analysis can be performed for any one of these Hamiltonians as well. Moreover for a given Hamiltonian, one may explore 
many different schemes for minimizing bending losses. Hence, the comparison of different Hamiltonians and different minimization schemes can be only 
a long-term project that goes clearly beyond the scope of the present manuscript.

%%%%%%%%%%%%%%%%%%%%%%%
\section*{Acknowledgements}
%%%%%%%%%%%%%%%%%%%%%%%
We acknowledge support from MSM6840770039 and SGS10/294/OHK4/3T/14.


\begin{thebibliography}{99}
\bibitem{remark10}
The optimization is performed numerically and agrees with \cite{UniProt} i.e.,   $\Omega\simeq 1.05 \Omega_0 N^{-1/6}$ .
\bibitem{remark1}
Stresses that remain after their original source (e.g., external forces, heat gradient) has been removed.
\bibitem{remark2} 
This is a universal ratio for a given chain i.e., it does not depend on the actual position of the bend. If necessary, 
it is straightforward to deduce the ratio $g/\Omega_{\alpha,\alpha\pm 1}$.
\bibitem{remark5} 
$\Omega_{\max}$ represents the upper limit on the achievable couplings within a physical realization of the chains, and is determined by physical or technological constraints.
\bibitem{remark6}
The effect of $\hat{{\mathscr V}}(\theta)$ can be treated perturbatively only  for $\kappa\ll 1$. Moreover, for the protocols under consideration and the entire regime of parameters, the couplings $\Omega_{1,2}$ and $\Omega_{N-1,N}$, cannot be treated perturbatively either.
\bibitem{remark3a} 
As a general remark we point out that for fixed $\kappa$ the effects of the perturbation on $Q_i$ and $S_i$ become more prominent if we bend the chain at regimes of smaller NN couplings. This implies that for both protocols 
the fastest drop of $Q_i$ and $S_i$ with $\kappa$ is observed for bends close to the ends (i.e., for $\alpha=2,N-1$), since in this region we have the smallest $\Omega_{\alpha,\alpha\pm 1}$. However, in the case of protocol 1, the variations in the behaviour of $Q_1(\kappa)$ and $S_1(\kappa)$ for $\alpha>2$ are not significant since $\Omega_{\alpha,\alpha\pm 1}=\Omega_0,\,\forall a\neq 2,N-1$.
\bibitem{remark4}
Detunings for $\kappa\gtrsim 0.7$ have to be taken with a pinch of salt, since they far exceed $\Omega_{\max}$ 
 and thus may not be realizable within certain physical platforms. In this case, non-optimal values of $\Delta_\alpha$ may be chosen to suppress partially bending losses.
\bibitem{remark8}
There are cases where the performance of an optimized bent chain 1, is slightly better than the performance of the corresponding unbent chain. Such improvements, however, are of the order of the tolerance used in our simulations (i.e., $\sim10^{-4}$), and can be neglected.
\end{thebibliography}
\end{document}